\documentclass[pre,aps,showpacs,byrevtex,amsmath,amssymb,twocolumn,preprintnumbers]{revtex4}
\usepackage{graphicx}% Include figure file
\usepackage{color}
\begin{document}
%\preprint{APS/123-QED}
%\preprint{PREPRINT}
\newcommand{\Cmm}{{C/m$^2$}}
\newcommand{\etal}{{\em et al.\/}}
\newcommand{\romd}{{\rm d}}

\title{On the regimes of charge reversal}
%in the restricted primitive model electrolyte}

%AUTHORS
%1st Author
\author{Felipe Jim\'enez-\'Angeles}
\email{fangeles@imp.mx}
%\affiliation{Departamento de
%F\'{\i}sica, Universidad Aut\'onoma Metropoloitana-Iztapalapa,
%Apartado Postal 55-334, 09340 D.F. M\'exico}
\affiliation{Programa de Ingenier\'{\i}a Molecular, Instituto
Mexicano del Petr\'oleo, L\'azaro C\'ardenas 152, 07730 M\'exico,
D. F., M\'exico}
%
%
%3rd Author
\author{Marcelo Lozada-Cassou}
\email{marcelo@imp.mx}
%\affiliation{Departamento de
%F\'{\i}sica, Universidad Aut\'onoma Metropoloitana-Iztapalapa,
%Apartado Postal 55-334, 09340 D.F. M\'exico}
\affiliation{Programa de Ingenier\'{\i}a Molecular, Instituto
Mexicano del Petr\'oleo, L\'azaro C\'ardenas 152, 07730 M\'exico,
D. F., M\'exico}
\date{\today{}}
\begin{abstract}
Charge reversal of the planar electrical double layer is studied
by means of a well known integral equations theory. By a numerical
analysis, a diagram is constructed with the onset points of charge
reversal in the space of the fundamental variables of the system.
%: the particles volume fraction ($\eta$), the ion-ion coupling
%($\xi$) and ion-surface coulomb coupling ($\gamma$) parameters.
%To the best of our knowledge, this is the first time a charge
%reversal diagram for the planar electric double layer is
%constructed.
%
Within this diagram two regimes of charge reversal are identified,
referred to as oscillatory and non oscillatory. We found that
these two regimes can be distinguished through a simple formula.
Furthermore, a symmetry between electrostatic and size
correlations in charge reversal is exhibited. The agreement of our
results with other theories and molecular simulations data is
discussed.
\end{abstract}
%\pacs{61.20.Qg,68.08.-p}
\maketitle

%
%
%HWG with the the MAIN TEXT
%
%%%%%
%\newpage

\section{Introduction}

Charged particles are naturally adsorbed onto an oppositely
charged surface, however, temperature prevents them to completely
condensate producing a diffuse ionic concentration profile known
as the electrical double layer (EDL). Intuitively one might expect
that the adsorbed counterions in the EDL are just the necessary to
compensate the surface's charge, however, for certain conditions,
they {overcompensate} it producing a surface {\em charge
reversal}. Hence, an inversion of the local electric field takes
place next to the first counterions layer and, consequently,
coions are attracted to form a second layer referred to as {\em
charge inversion} of the EDL.

{Charge reversal} (CR) and {charge inversion} (CI) have motivated
a large number of studies in the past
\cite{LevinRVW_2002,gelbart00,attard_1995,
jimenez_2001,tanaka_2001,terao_2001,terao_2002,Jimenez2,Jimenez3,
messina_PRL2000,messina_PRE_01,messinaEPL_2002,shklovskiiL2000,kjellander96,kjellanderEChA96}.
These effects have been observed in the formation of
self-assembled polyelectrolyte layers on a charged substrate
\cite{decher97}, self-assembled DNA-lipid membrane complexes
\cite{raedler_sci2} and anomalous macroions adsorption on Lagmuir
films \cite{rondelez_1998}. Furthermore, CR has been associated
with reverse of the electrophoretic mobility of charged colloids
\cite{lozada99,lozada_2001,hidalgoRVW_2002,Tanaka2004} and
polyelectrolytes \cite{HsiaoPRL06} as well as with attraction
between like charged macroparticles
\cite{lozada90a,kekicheff95,jimenezJCP06} and DNA condensation
\cite{BestemanNat07}.

CR and CI manifest as alternated layers of coions and counterions
in the neighborhood of a charged interface, i.e., imply spatial
oscillations of the inhomogeneous charge distribution profiles.
For bulk electrolytes, the oscillatory behavior of the pair
correlation functions was predicted by Stillinger and Lovett
\cite{stillinger68} as a consequence of their second moment
condition.
%which implies that oscillations of
%the pair correlation function will occur when a relationship
%between the ionic excluded volume and Coulombic coupling is
%satisfied.
The oscillations of the inhomogeneous charge density profiles have
been observed in studies where the restricted primitive model
(RPM) electrolyte is used to model the diffuse EDL
\cite{megen80,outhwaite_1981,lozada82,Torrie_1982,gonzales_1985,attard_1996,kjellander_1998,kjellanderEChA96}.
%In the restricted primitive model (RPM) electrolyte the ions are considered as a
%mixture of equally sized charged hard spheres embedded in a
%continuous medium.
For the EDL produced by a RPM electrolyte at a planar interface,
two regimes of CR have been distinguished
\cite{attard_1996,kjellander_1998}, i.e., (i) the oscillatory and
(ii) non oscillatory. These regimes are named according to the
behavior of the electrolyte pair correlation functions in bulk. In
the former regime, since the bulk pair correlation functions
oscillate, any disturbance caused by an external electric field is
propagated in the same way, therefore, {CR} occurs for any {\em
nonzero} surface charge density. In the second regime, a
sufficiently high surface charge density is necessary to produce
CR. In both regimes, short range and long range correlations
correlations play an important role in the surface CR
\cite{kjellanderEChA96,jimenez_2001}.

For some cases, the onset points of oscillations of the bulk ionic
pair correlation functions \cite{kjellanderJCP95} and EDL-CR
\cite{jimenez_2001,terao_2001,messina_EPL_00} have been computed,
however, a complete picture of both phenomena and their
interrelationship is still missing. To fill this void, we
construct a CR diagram for the planar EDL in terms of its
fundamental variables, i.e., the particles volume fraction
($\eta$), the ion-ion ($\xi$) and ion-wall ($\gamma$) Coulomb
coupling. We also analyze the dependence of CR on each variable.
Hence, the reminder of the paper is organized as follows: in
section \ref{theory} we describe the model and theory. Section
\ref{results} contains the results and their discussion, and,
finally, some conclusions are given in the closing part.

%Here, the EDL of a planar wall next to a restricted primitive
%model electrolyte is studied by means of the hypernetted chain/
%mean spherical approximation (HNC/MSA) theory for inhomogeneous
%fluids.
%%
%By numerically scanning the onset points of charge reversal, a ``
%phase diagram" is constructed in terms of the system's fundamental
%variables, i.e., the particle's volume fraction ($\eta$), the
%ion-ion ($\xi$) and ion-wall ($\gamma$) coulomb coupling. Within
%this diagram the regimes of charge reversal are distinguished by
%means of a simple formula.
%As the main result of this
%study, a criterion to distinguish the regimes of charge reversal
%is given.

%\begin{figure}
%\includegraphics[width=8.0cm]{fig1.eps}
%\caption{ (a)A diagrammatic representation of the many body {\em
%transversal} effective force on a single particle in the EDL. (b)
%A schematic representation of a {\em lateral} many body effect on
%a adsorbed particle. The arrows represent the repulsive
%electrostatic forces and collisions.} \label{mechanistic}
%\end{figure}

\section{Theory}
\label{theory}

Particles and fields are equivalent since both are defined
thorough their interaction with other particles and fields. This
simple fact has been used in physics of inhomogeneous fluids to
derive integral equations theories \cite{henderson92a}, i.e., in a
nonuniform fluid the external field is taken as one more species
of the fluid. Hence, one can use standard integral equations
theories for uniform fluids based on the Ornstein-Zernike
equation.

In the Ornstein-Zernike equation for inhomogeneous fluids, the
total correlation function between an external field and a
particle of the fluid, $h_{\alpha i}({\bf r}_{21})$, is decomposed
as the sum of direct and indirect correlations, i. e.,
\begin{equation}
h_{\alpha i}({\bf r}_{21})=c_{\alpha i}({\bf
r}_{21})+\sum_{k=1}^n\int{\rho_m h_{\alpha m}({\bf r}_{23})c_{m
i}({\bf r}_{31})}d{\bf r}_{3} \label{eq:oz}
\end{equation}
where $c_{\alpha i}({\bf r}_{21})$ represents the direct
correlation function between the external field and a particle of
species $i$ located at ${\bf r}_{21}$ and $c_{m i}({\bf r}_{31})$
is the direct correlation function between any pair of particles,
of species $m$ and $i$, in the fluid. In this notation the
subindex $\alpha$ refers to the external field. The second term in
Eq.~\eqref{eq:oz} represents the indirect correlation between the
external field with a particle of species $i$, carried out through
all the particles of species $m$ in the fluid. The total
correlation functions are related to the particles distribution
functions, $g_{\alpha i}({\bf r}_{21})$, by $h_{\alpha i}({\bf
r}_{21}) = g_{\alpha i}({\bf r}_{21})-1$.

The Ornstein-Zernike equation requires extra relationships,
referred to as closures,  between $h_{\alpha i}({\bf r}_{21})$ and
$c_{\alpha i}({\bf r}_{21})$ to be solved. Here, we apply the
hypernetted chain/mean spherical approximation (HNC/MSA), which
implies that  $c_{\alpha i}({\bf r}_{21})$ is approximated by the
hypernetted chain (HNC) closure, whereas $c_{m i}({\bf r}_{31})$
is taken through the mean spherical approximation (MSA). This
approach has been applied for modelling inhomogeneous and confined
electrolytes in many geometries and, in general, it has shown good
agreement with computer simulations
\cite{degreve_1993,lozadaPRL_1997,jimenez_2001}.

\subsection{The surface and fluid models}

\begin{figure}
\includegraphics[width=6.0cm]{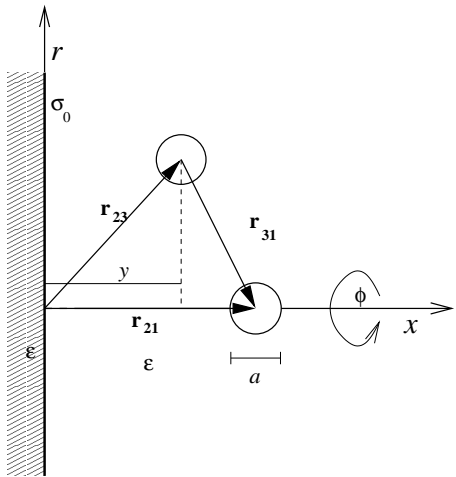}
\caption{Set up of the system under study.} \label{setup}
\end{figure}

We consider an electrolyte solution contiguous to a planar wall
which has a uniform surface charge density $\sigma_0$ and
dielectric constant $\epsilon_w$ (See Fig.~\ref{setup}). In the
restricted primitive model (RPM) electrolyte, the ions are
considered as hard spheres of diameter $a$, with a centered point
charge $q_i=ez_{i}$($e$ is the proton's charge and $z_{i}$ is the
ionic valence), and at concentration $\rho_i$. The ions are
embedded in a structureless solvent, considered only through a
uniform medium of dielectric constant $\epsilon$ and, for
simplicity, $\epsilon_w=\epsilon$. Within this model, a high
particles volume fraction (large $a$ and/or high $\rho_i$) could
be a simplistic way of taking into account solvent effects, such
as solvation or excluded volume. In this regard, some recent works
have considered the solvent effect through more refined models
\cite{Gavryushov2007,Lamperski2007}.

In planar geometry, $g_{\alpha i}({\bf r}_{21})$ depends only of
the perpendicular distance to the surface, $x$, therefore
hereinafter we will simply write $g_i(x)\equiv g_{\alpha i}({\bf
r}_{21})$ and $h_m(y)\equiv h_{\alpha m}({\bf r}_{23})$, where the
subindex $\alpha$ has been omitted. The charge on the wall is
compensated by an excess of charge in the fluid given by the
charge density profile, $\rho_{el}(x)$, such that

\begin{equation}
-\sigma_0=\int_{a/2}^{\infty}\rho_{el}(x)dx
\end{equation}
%%%
The charge density profile gives the structure of the electric
double layer (EDL) and it is expressed in terms of the ions
distribution functions, $g_i(x)$, as

\begin{equation}
\rho_{el}(x)=\sum_{i=1}^2z_ie \rho_i g_i(x)
\end{equation}
%%%

The surface-ion direct interaction potential is written as
$u_{\alpha i}(x)=u^{\rm hs}_{\alpha i}(x)+u^{\rm el}_{\alpha
i}(x)$, where the first term is the ion-surface hard-core
interaction potential which considers that ions can not penetrate
or deform the surface. The second term is the electrostatic
contribution, given by

\begin{equation}
u_{\alpha i}^{\rm el}(x)= -\frac{2 \pi}{\epsilon}q_i\sigma_0
(x-x_\infty) \label{potential}
\end{equation}
being $x_\infty$ a reference point.

The $i$ species concentration profile, $g_i(x)$, is related to the
ion-wall potential of mean force, $w_i(x)$, by

\begin{equation}
{g_i(x)= \exp \{-\beta w_i(x)\}}
\end{equation}
from which the ion-wall effective mean force, $f_{xi}(x)$, is
obtained by
\begin{equation}
f_{xi}(x)=-\frac{\partial w_i(x)}{\partial x}= k_B T
\frac{\partial\ln g_i(x)}{\partial x}
\end{equation}
From this relationship one can infer that a negative (positive)
slope of $g_i(x)$ implies an attractive (repulsive) behavior of
$f_{xi}(x)$.

The computation of $g_i(x)$ requires to solve Eq.~\eqref{eq:oz}.
As we pointed out above, it is solved by approximating
$c_i(x)\equiv c_{\alpha i}({\bf r}_{21})$ by the hypernetted chain
closure in Eq.~\eqref{eq:oz}, i.e.,
\begin{equation}
c_i(x)=-\ln{g_i(x)}-\beta u_i(x)+h_i(x) \label{hnc}
\end{equation}
%%%
Thus, by combining Eqs.~\eqref{hnc} and \eqref{eq:oz} we get
%%%
\begin{equation}
g_{i}(x)=\exp \left\{-\beta u_{\alpha i}(x)
+\sum_{m=1}^{2}\rho_{m}\int h_{m}(y) c_{mi}^{MSA}(s)\ d{\bf
r}_3\right\} \label{eq:hnc}\end{equation}
%%
%%
%where, $g_{i}(x)$ is the ions distribution function for the
%species $i$ at a distance $x$ from the wall, $h_{m}(y)$ $ \equiv
%g_{m}(y)-1$ is the total correlation function for species $m$,
%$\rho_{j}$ is the bulk number density of that ionic component and
In addition, in Eq.~\eqref{eq:hnc}, we used the bulk MSA direct
correlation functions between two ions of species $i$ and $m$,
i.e., $c_{mi}^{MSA}(s)$, with $s$$\equiv$$|{\bf r}_{31}|$ their
relative separation distance. As usual, $\beta =1/(k_{B}T)$, where
$k_{B}$ is the Boltzmann constant and $T$ is the absolute
temperature. The explicit form of $c_{mi}^{MSA}(s)$ for the RPM
electrolyte can be written as

\begin{equation}
c_{mi}^{MSA}(s)=c^{\rm hs}(s)+q_m q_i c^{\rm sr}(s)-\frac{\beta
q_m q_i}{\epsilon s}
\end{equation}
where $c^{\rm hs}(s)$ and $c^{\rm sr}(s)$ are short range
functions, being equal to zero for $s>a$. By integrating
Eq.~(\ref{eq:hnc}) in cylindrical coordinates, after some algebra
it is written as

\begin{eqnarray}
g_{i}(x)&=& \exp\left\{\frac{4\pi \beta}{\epsilon}z_ie
\sigma_0x + 2\pi\int_{a/2}^{\infty}h_{s}(y) K(x,y)dy \nonumber \right.\\
&+& J(x)+  2\pi z_i\int_{a/2}^{\infty}h_{d}(y)L(x,y)dy \label{rpm-hnc-msa_placa4} \\
&+&\left. \frac{2\pi\beta e^2
z_i}{\epsilon}\int_{a/2}^{\infty}h_{d}(y)[x+y+|x-y|]dy\right\}\nonumber
\end{eqnarray}
with $i=1,2$ and defining $h_{s}(y)\equiv\sum_{m=1}^2 \rho_m
h_{m}(y)$, $h_{d}(y)\equiv\sum_{m=1}^2  z_m \rho_m h_{m}(y)$ and
\begin{eqnarray*}
K(x,y)&\equiv&\int_{|x-y|}^{\infty}c_{s}(s)sds\\
L(x,y)&\equiv&\int_{|x-y|}^{\infty}c_{d}^{sr}(s)sds\\
J(x)&\equiv&-2\pi\rho_T\int_{x-a}^{a/2}K(x,y)dy
\end{eqnarray*}
where $\rho_T=\sum_{i=1}^2\rho_i$ and it has been used that
$h_{j}(x)=-1$ for $x\le a/2$. The HNC/MSA integral equations,
Eq.~(\ref{rpm-hnc-msa_placa4}), can be conveniently written as

\begin{equation}
g_{i} (x) = \exp \{-\beta w_i(x)\}=\exp \left\{-\beta q_{i}
\psi(x) + w_{i}^{\rm sr}(x)\right\} \label{hnc:msa2}
\end{equation}
where $\psi(x)$ is the mean electrostatic potential, expressed as,
\begin{equation}
\psi(x)=-\frac{4 \pi}{\epsilon}\sigma_0 x-
\frac{2\pi}{\epsilon}\int_{a/2}^{\infty} {\rho_{\rm
el}(y)}[x+y+|x-y|]dx
\end{equation}
and $w_i^{\rm sr}(x)$ is a short range potential, given by,

\begin{eqnarray}
w_i^{\rm sr}(x)&=&-2\pi\int_{a/2}^{\infty}h_{s}(y) K(x,y)dy - J(x)
-2\pi z_i\int_{a/2}^{\infty}h_{d}(y) L(x,y)dy \label{wsr}
\end{eqnarray}
In Eq.~\eqref{wsr} size correlations are included through the
first two terms, meanwhile, the last term includes electrostatic
short range correlations. Thus, in the case of $q_i=0$,
$\psi(x)=0$ and only the size correlations terms are included in
Eq.~(\ref{hnc:msa2}), hence, the model reduces to the
inhomogeneous hard-sphere fluid.
On the other hand, in the point ions limit ($a=0$), $w^{\rm
sr}_i(x)=0$ and Eq.~(\ref{hnc:msa2}) becomes

\begin{eqnarray}
g_{i} (x) &=&\exp \left\{-\beta q_{i} \psi(x)\right\} =\exp
\left\{\frac{4 \pi}{\epsilon}\sigma_0\beta q_i x\right. \nonumber \\
& + &\left. \frac{2\pi}{\epsilon}\beta q_i\int_{a/2}^{\infty}
\rho_{\rm el}(y)[x+y+|x-y|]dy \right\} \label{placa:PB}
\end{eqnarray}
which is the integral version of the Poisson-Boltzmann (PB)
equation. In Eq. (\ref{placa:PB}) a closest approach distance of
the ions to the wall, $a/2$, is considered. PB equation for
size-symmetric electrolytes does not predict CR
\cite{jimenez_2001} since both, size and electrostatic short range
correlations are neglected. However, it should be mentioned that
models based on the PB equation may predict charge reversal when
size correlations are considered artificially, i.e., by assigning
to cations and anions unequal closest approach distances to the
wall \cite{Eloy2JCIS}.

Eqs. (\ref{hnc:msa2}) and (\ref{placa:PB}) depend on several
parameters, i.e., on $\rho_i$, $a$, $T$, $\epsilon$, $z_ie$,
$\sigma_0$. To simplify the analysis below we introduce the three
fundamental parameters that describe the planar EDL.

\subsection{The fundamental parameters}

The thermodynamical properties of the symmetric RPM electrolyte
can be described exclusively in terms of the following parameters
\cite{stillinger68}

\begin{equation}
\xi \equiv \frac{q_i^2\beta}{a\epsilon}
\end{equation}
%%%
%%%
%%%
\begin{equation}
\eta\equiv\frac{\pi}{6}\rho_Ta^3
\end{equation}
$\xi$ accounts for strength of the ion-ion electrostatic
interaction and is referred to as the ion-ion Coulomb coupling.
The second parameter, $\eta$, is the particles volume fraction.

The ion-wall direct electrostatic interaction at the wall's
surface is given by
\begin{equation}U_{i}=q_i u_{\alpha i}\left(\frac{a}{2}\right)
   =\frac{2\pi
   q_i\sigma_0}{\epsilon}\left(x_{\infty}-\frac{a}{2}\right)
\end{equation}
thus, we define a third parameter
%%%
\begin{equation}
\gamma_i \equiv \frac{2\pi\beta}{\epsilon} q_i \sigma_0 a
\label{pram:gamma}
\end{equation}
which quantifies the ion-wall electrostatic interaction. We also
define $\gamma\equiv|\gamma_i|$. It is an easy task to show that
total potential energy and the partition function of the system
(charged wall + electrolyte) can be written in terms of these
dimensionless variables.

%The more negative the value of $U_i$, particles adsorption is
%energetically more favorable. Several important effects are beyond
%this effect.

In terms of the $\gamma_i$ parameter, Eq.(\ref{potential}) can be
written as
%%%
\begin{equation}
-\beta u_{\alpha i}(X)= \frac{1}{2}\gamma_i (X-X_\infty)
\end{equation}
with $X\equiv 2x/a$ the dimensionless distance. Hence,
Eq.~(\ref{rpm-hnc-msa_placa4}) is written as

\begin{eqnarray}
g_i(X)&=&\exp \left\{\gamma_i X+ \int_1^\infty \hat{h}_{\alpha s}(Y)\hat{K}(X,Y)d{Y}\nonumber \right.\\
&+&\hat{J}(X)+ \hat{z_i}\xi\int_1^\infty \hat{h}_{\alpha d}(Y)
\hat{L}(X,Y)d{Y} \label{rpm-hnc-msa_placa5} \\  &+&
\left.3\hat{z_i}\xi \int_1^\infty \hat{h}_{\alpha
d}(Y)[X+Y+|X-Y|]d{Y}\right\}\nonumber
\end{eqnarray}
where $\hat{z}_j\equiv q_j/q$ and we have defined
\begin{eqnarray*}
\hat{h}_{\alpha s}(X)=\sum_{m=1}^n \eta_m h_{\alpha m}(X) \\
\hat{h}_{\alpha d}(X)=\sum_{m=1}^n  \hat{z}_m \eta_m h_{\alpha m}(X)\\
\end{eqnarray*}
The expressions for $\hat{K}(X,Y)$, $\hat{L}(X,Y)$ and
$\hat{J}(X)$ are given below. For $X-2\le Y \le X+2$,
\begin{eqnarray*}
\hat{K}(X,Y)&=&\frac{3}{4}c_1[(X-Y)^2-4]+\frac{3}{2}\eta c_2[|X-Y|^3-8]\\
&+ &\frac{3}{160}\eta c_1[|X-Y|^5-32] \\
\hat{L}(X,Y)&=&3 [2-|X-Y|]- \frac{3}{2}\frac{\Gamma
a}{(1+\Gamma a)}[4-(X-Y)^2] \\
&+&\frac{1}{4}\left(\frac{\Gamma a }{1+\Gamma
a}\right)^2[8-|X-Y|^3]
\end{eqnarray*}
otherwise $\hat{K}(X,Y)=\hat{L}(X,Y)=0$. For $1\le X \le 3$
\begin{eqnarray*}
\hat{J}(X)&=&\frac{1}{4}\eta c_1[\hat{p}^3-12\hat{p}+16]
+\frac{3}{8}\eta^2 c_2[\hat{p}^4-32\hat{p}+48] \\ \nonumber
&+&\frac{1}{320}\eta^2c_1[\hat{p}^6-192\hat{p}+320]
\end{eqnarray*}
otherwise $\hat{J}(X)=0$, with $\hat{p}\equiv X-1$.
Eq.~(\ref{rpm-hnc-msa_placa5}) is solved using a finite element
technique. Details of the numerical method are given elsewhere
\cite{lozadaJCompP}. We wish to point out that the reduced
concentration profiles, $g_i(X)$, depend only on the fundamental
parameters $\xi$, $\eta$ and $\gamma$, for any combination of
$\rho_i$, $a$, $T$, $\epsilon$, $z_ie$, $\sigma_0$. That is, two
systems are equivalent provided that $\eta$, $\xi$ and $\gamma$
are the same for both systems.

\section{Results and discussion}
\label{results}

The local electric field, $E(x)$, is useful to identify and
quantify charge reversal. By Gauss' law it is given by

\begin{equation}
E(x)=\frac{4\pi}{\epsilon}[\sigma_0+\sigma'(x)] \label{eq:field}
\end{equation}
being $\sigma'(x)$ the surface charge density accumulated between
two planes, parallel to the surface at $a/2$ and $x$,
respectively, given by
\begin{equation}
\sigma'(x)=\int_{a/2}^x\rho_{\rm el}(y)dy \label{sigma_profile}
\end{equation}

The electroneutrality condition implies that ${\displaystyle
\lim_{x\rightarrow\infty}\sigma'(x)=-\sigma_0}$, such that the
electric field produced by the plate is completely screened at
infinity.
In order to evaluate {charge reversal} (CR) we define the
dimensionless function
\begin{equation}
\sigma(x)\equiv-\frac{\sigma_0+\sigma'(x)}{\sigma_0}
\end{equation}
such that CR occurs in an interval $[x_1 , x_2]$ where
$\sigma'(x)$ overcompensates $\sigma_0$, or equivalently,
$\sigma(x)>0$. Conveniently we also define $\sigma^\ast$ as the
absolute maximum of $\sigma(x)$, i.e.,

\begin{equation}
\sigma^\ast\equiv \mbox{max}\{\sigma(x):x\in [a/2,\infty)\}
\end{equation}
which, from Eq. (\ref{sigma_profile}), is written as

\begin{equation}
\sigma^\ast=-1-\frac{1}{\sigma_0}\int_{a/2}^{x_{max}}\rho_{el}(x)dx
\label{eq:maxCR2}
\end{equation}
being $x_{max}$ the location of the absolute maximum of
$\sigma(x)$. For the symmetric RPM electrolyte, which is the case
considered here, $z_+$$=-z_-$$=z$ and $\rho_+$$=\rho_-$$=\rho_s$.
Furthermore, it can be readily shown that the extreme values of
$\sigma(x)$ are located at a set of points, $x_i$, where
$\rho_{el}(x_{i})=\sum_{i=1}^2z_ie\rho_i g_i(x_i)=0$, implying
$g_+(x_{i})=g_-(x_{i})$. It will be shown that the absolute
maximum of $\sigma(x)$, $\sigma^\ast$, is the closest one to the
wall. Thus, the plane at $x=x_{max}$ defines the boundary of the
adsorbed counterions layer and $\sigma^\ast$ quantifies the rate
of charge overcompensation.

%(ii) When $\eta$ and $\xi$ are {\em low} a successive increasing
%of $\gamma$ eventually produces {charge reversal}.

%occurs and always increases by increasing $\eta$ and $\xi$
%From a primary exploration in terms of $\eta$, $\xi$ and $\gamma$,
%we found the following: (i) for moderately low values of the
%$\gamma$ parameter (typically $\gamma\ll 1$)

\begin{figure}
\mbox{\put(68,65){\includegraphics[width = 4.3
cm]{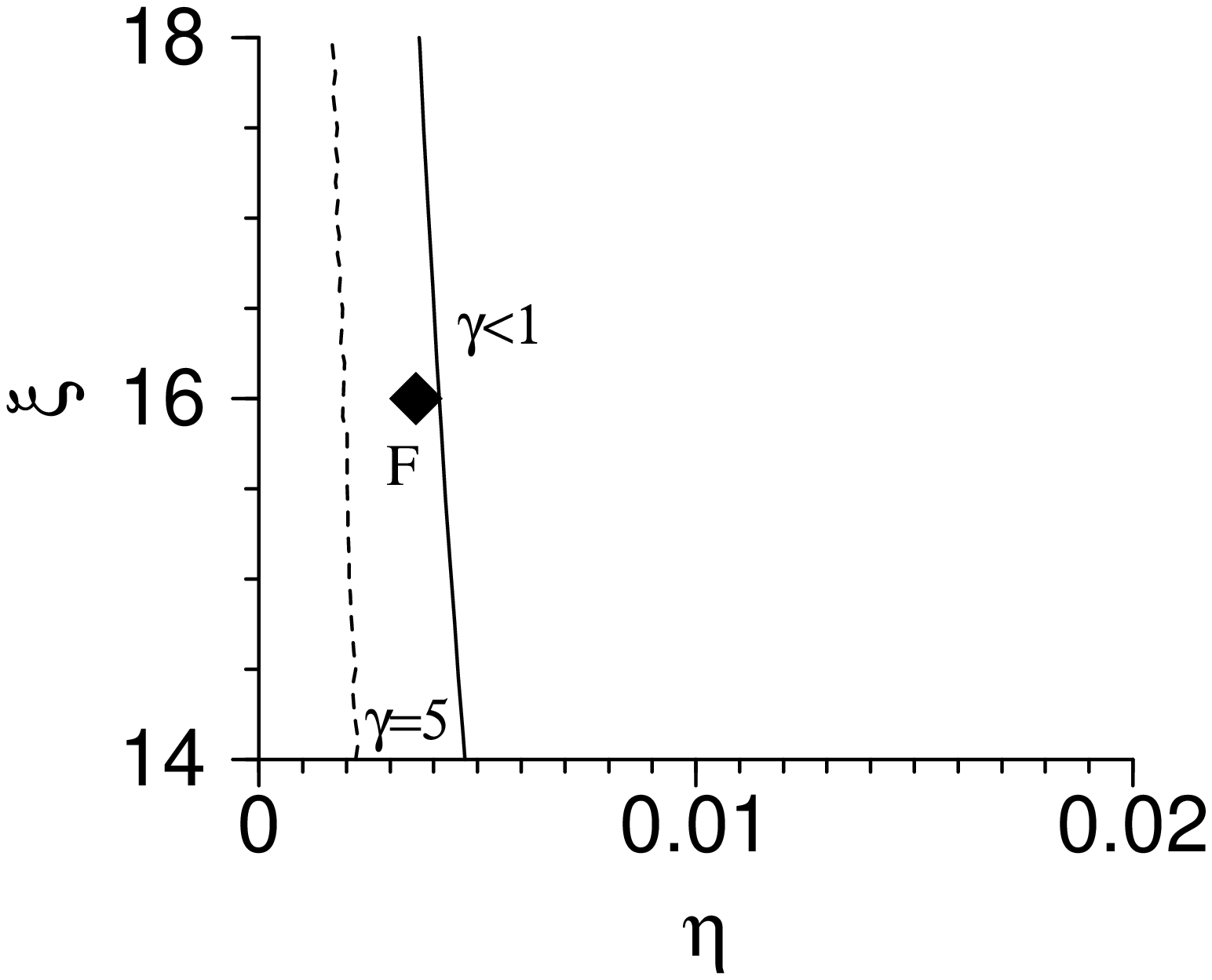}}\includegraphics[width = 7.0 cm]{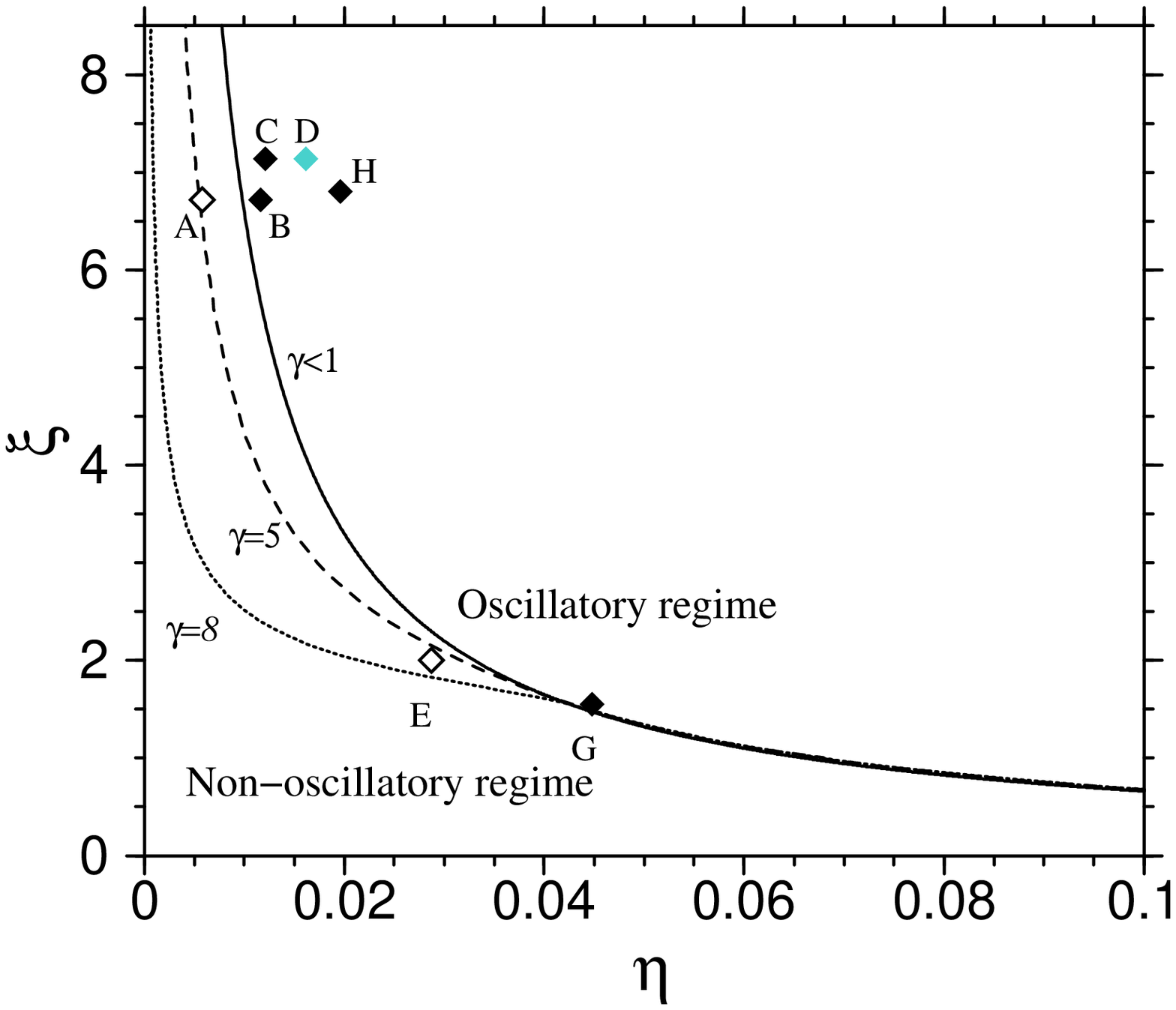}}
\caption{Charge reversal (CR) diagram in the $\xi$-$\eta$ plane
with three CR curves: $\gamma\lesssim 1$ (solid line), $\gamma= 5$
(dashed line), and $\gamma= 8$ (dotted line). The systems
described in Table I are plotted within the diagram: Black symbols
are onset points of CR, whereas white and light symbols mean the
absence and occurrence of CR, respectively. The inset is an
amplification for high values of $\xi$. For the onset points of CR
from molecular simulation (b,c,f) the numerical error is
approximately the symbol size.} \label{diagram1}
\end{figure}

We constructed a {\em charge reversal diagram}\cite{noteCR} in the
$\eta$-$\xi$-$\gamma$ space in the following way: For a given
constant value of $\gamma$ and starting from values of $\xi$ and
$\eta$ close to zero, both parameters were progressively raised
until the condition $\sigma^*\gtrsim 0$ was reached (for numerical
purposes we set $\sigma^*\gtrsim 10^{-5}$). Hence, by repeating
this procedure for different values of $\gamma$ the full space was
analyzed. The onset of points of CR ($\eta^t$,$\xi^t$) define a
curve referred to as the {\em charge reversal curve}.
Thus, the main result of this work is presented in
Fig.~\ref{diagram1}: the charge reversal diagram  for the planar
EDL. For clarity only three $\gamma$-dependent CR curves are
plotted. For each value of $\gamma$, below the curve charge
reversal does not occur whereas it does above it. It should be
mentioned that $\sigma^\ast$ continues increasing for higher
values of either parameter, $\xi$ or $\eta$.
Interestingly, for $0<\gamma\lesssim 1$ all curves converge to the
solid line. The dashed and dotted lines are for $\gamma=5$ and
$\gamma=8$, respectively. Curiously, the three CR curves converge
for $\xi \lesssim 1.7$. In general, as $\gamma$ increases the CR
curves are shifted to the left. It should be mentioned that our
diagram covers a broad range of the $\eta$-$\xi$-$\gamma$ space,
including many cases of practical interest.
%
%To the best of our knowledge this is the first time a CR diagram
%is constructed for the planar EDL, notwithstanding,  Nguyen and
%Shklovskii \cite{shklovskiiJCP01} have suggested a
%phenomenological phase diagram for charge inversion and
%complexation for DNA a solution with charged spheres.

%%%%%%%%%%%%%%
%%%
%%%%%%%%%%%%%%

%%%
\begin{table}
  \centering
  \begin{tabular}{c c c c c c l}
  % after \\: \hline or \cline{col1-col2} \cline{col3-col4} ...
  \hline
  \text{Label}& $\eta$ & $\xi$ &\text{$a$[\AA]} & $\rho_s$ & $z$ & Method \\
  \hline \hline
  a& 5.81 $\times 10^{-3}$& 6.72 &4.25 & 0.12 & 2 &\text{MD \cite{jimenez_2001}} \\
  b& 1.162 $\times 10^{-2}$& 6.72 &4.25 & 0.24 & 2 & \text{MD \cite{jimenez_2001}}\\
  c&1.21$\times 10^{-2}$&7.14   &4.0 &0.3 &2 &   \text{MC \cite{terao_2001}} \\
  d&1.61$\times 10^{-2}$& 7.14 &4.0 &0.4 &2 & \text{MC \cite{terao_2001}} \\
  e& 3.58 $\times 10^{-3}$& 16  &1.785&  1.0   & 2 & \text{MD \cite{messina_EPL_00}}\\
  f&2.869 $\times 10^{-2}$& 16  &3.57&  1.0   & 2 & \text{MD \cite{messina_EPL_00}}\\
  g& $4.47\times 10^{-2}$& 1.552&4.6& 0.7293&  1 &    \text{DIT \cite{kjellanderJCP95}}\\
  h&1.963$\times 10^{-2}$& 6.8 &4.2&0.420& 2 & \text{DIT \cite{kjellanderJCP95}}\\
\hline
\end{tabular}
\caption{Numerical values ($\eta$ and $\xi$ ) for the points (a-h)
in Fig.~\ref{diagram1}, with their corresponding values of ionic
diameter, $a$, electrolyte concentration, $\rho_s$, and valence,
$z$. In all cases $l_B=\frac{\beta e^2}{\epsilon}\approx 7.14$\AA.
Method refers to dressed ion theory (DIT), molecular dynamics (MD)
and Monte Carlo (MC) simulations.}\label{Table}
\end{table}

%We fitted a function to  and
A least squares fitting in a log-log scale was carried out for the
data composing the solid line of Fig.~\ref{diagram1}.
Surprisingly, we found that they satisfy the simple relationship
\cite{numerical} $\eta \xi=0.0662$, or equivalently $24 \xi
{\eta}=\kappa^2 a^2 \approx 1.6$ , with
$\kappa^2=\frac{4\pi\beta}{\epsilon}\sum_{i=1}^2q_i^2 \rho_i$, the
inverse Debye length. This curve splits the $\eta$-$\xi$ plane in
two regimes: Regime I ($\xi\eta>0.0662$) and regime II
($\xi\eta<0.0662$). In regime I, {CR} occurs for any value of
$\gamma$ different from zero, that is, CR is a property of the
fluid due to the fact that the bulk ion-ion pair correlation
function is oscillatory. This implies that any perturbation in the
fluid (independently of the external field strength) is propagated
in an oscillatory manner, producing alternated layers of
counterions and coions next to the wall, hence, CR and CI.
Although not shown, we found that this is valid for different
geometries of the external field. In regime II, the ion-ion pair
correlation function is monotonic, hence, the occurrence of CR
depends on the external field. According to the behavior of the
pair correlation function in bulk, regimes I and II are referred
to as oscillatory and non-oscillatory, respectively.

The onset points of CR for some EDL systems, from molecular
dynamics (b,f) and Monte Carlo (c) simulations, are plotted in the
charge reversal diagram of Fig.~\ref{diagram1}. We also plotted
the onset of oscillations of the ionic pair correlation function
predicted by the dressed ion theory (DIT) (g,h) (See table I). It
can be seen that our prediction ($\kappa a$$\approx$ 1.260, solid
line) well agrees with the onset of oscillations of the pair
correlation function predicted by DIT \cite{kjellanderJCP95} (
$\kappa a$$\approx$ 1.293), for a particular case of a monovalent
RPM electrolyte (g). Our prediction is also in close agreement
with the onset points of CR predicted by molecular dynamics (b,f)
and Monte Carlo (c) simulations for several EDL systems of
divalent electrolytes. For divalent electrolytes, our prediction
is in better agreement with simulation data than DIT (h).

\begin{figure}
{\includegraphics[width=7.0cm]{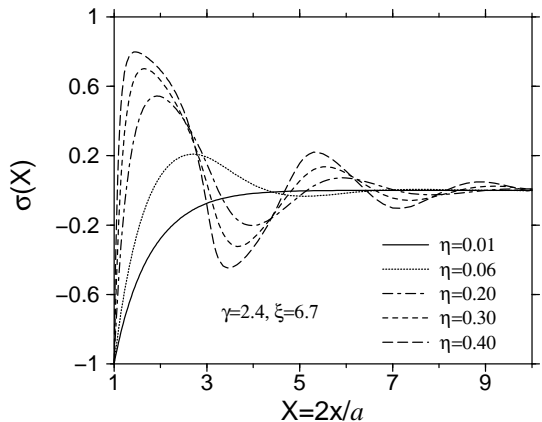}} \caption{$\sigma$(X)
for successively increasing values of $\eta$ and constant values
of $\xi=6.7$, and $\gamma=2.4$.} \label{coulomb_coupling}
\end{figure}

In Fig.~\ref{coulomb_coupling} $\sigma$ is plotted, as a function
of the dimensionless distance $X=2x/q$, for $\xi=6.72$,
$\gamma=2.4$ and $\eta=0.01,0.060.20,0.30,0.40$. Note that the
number and amplitude of oscillations increase as $\eta$ increases.
More importantly, however, is the fact that the absolute maximum
of $\sigma(X)$, $\sigma^\ast$, increases by increasing $\eta$. A
similar behavior is observed (not shown) if $\eta$ is constant and
$\xi$ is increased. As we mentioned above, in
Fig.~\ref{coulomb_coupling}, $\sigma^\ast$ is the closest maximum
to the wall.

\begin{figure}
{\includegraphics[width=7.0cm]{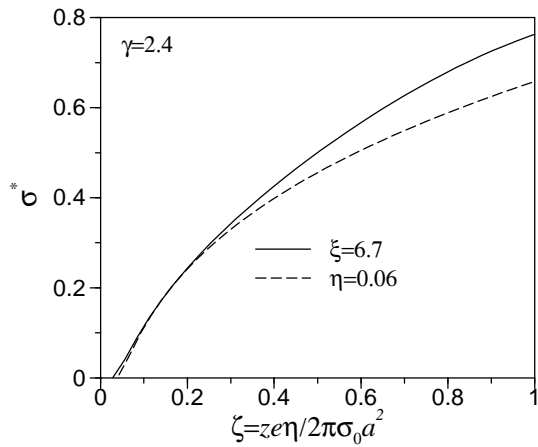}} \caption{$\sigma^{\ast}$
as a function of $\zeta=\xi\eta/\gamma=ze \eta/2\pi \sigma_0 a^2$,
for $\gamma=2.4$. The continuous line correspond to $\xi=6.7$ and
increasing values of $\eta$, whereas the dashed line is for
$\eta=0.06$ and increasing values of $\xi$.}
\label{coulomb_coupling2}
\end{figure}

CR is not only governed by the ion-wall direct interaction, but
above all, it is a many body effect in which long range and short
range correlations participate. In Fig.~\ref{coulomb_coupling2},
$\sigma^*$ is plotted as a function of
\begin{equation}
\zeta \equiv \frac{\xi
\eta}{\gamma} = \frac{ze \eta}{2\pi \sigma_0 a^2}
\end{equation}
For the solid line the particles volume fraction $\eta$ is raised
and the Coulomb coupling parameters are kept constant at
$\xi=6.72$ and $\gamma=2.4$. For the dashed line $\eta=0.06$,
$\gamma=2.4$, and $\xi$ increases. Surprisingly, in both cases
$\sigma^*$ increases monotonically with $\zeta$ in a qualitatively
similar way. In fact, for low values of $\zeta$ both curves are
overlapped. {\em In this sense we refer to electrostatic and size
correlations as symmetric}. This behavior, a priori, was not
expected since $\eta$ and $\xi$ quantify contributions of
different nature, i.e., size correlations ($\eta$) and
electrostatic correlations ($\xi$), respectively. The symmetry is
also exhibited in the charge reversal curves of
Fig.~\ref{diagram1}, and particulary for $\gamma$$\lesssim$1
($\xi$$\eta$=0.0662): when $\eta$ ($\xi$) is low, a higher value
of $\xi$ ($\eta$) is needed to produce charge reversal.

%an effective particle-wall attraction when particles are near the wall and
%Mechanistically seen, the particle-particle repulsion on
From a fundamental point of view, the extra adsorption of
particles to the wall, i.e., CR, implies a gain of entropy respect
to the bulk \cite{Jimenez2,Jimenez3}, thus, adsorption is favored.
A simple way of understanding this, is through an analogy with the
hard sphere model fluid next to a hard wall. In that case, size
correlations induce adsorption of particles to the wall, which
increases as the particle's size or concentration increase, i.e.,
as $\eta$ increases. In the HNC/MSA theory for a RPM electrolyte,
electrostatic and size correlations are taken into account.
Perhaps their symmetry (Figs.~\ref{diagram1} and
\ref{coulomb_coupling2}), can be explained by an effective
excluded volume that mutual electrostatic repulsion between like
charged ions confers them. Therefore, the higher value of $\xi$,
the higher the particles repulsion and the higher effective
excluded volume, hence, the higher CR.

\begin{figure}
\includegraphics[width = 7.0 cm]{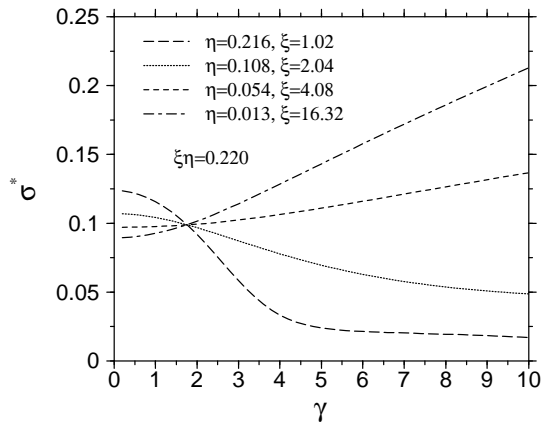}
\caption{$\sigma^{\ast}$ as a function of
$\gamma=\frac{2\pi\beta}{\epsilon}q_i\sigma_0 a$, with
$\xi\eta=0.220$, in all cases. An increasing of $\gamma$ is
equivalent to an increasing of $\sigma_0$.} \label{gamma}
\end{figure}

In Fig.~\ref{gamma}, we evaluate the effect of $\gamma$ on
$\sigma^\ast$. For each curve different values of $\eta$ and $\xi$
are combined, but a fixed value of $\xi \eta=0.220$ is set. The
highest considered value of the particles volume fraction is
$\eta=0.216$ with $\xi=1.02$. For this particular case, the number
of adsorbed counterions per unit area, $n^{cou}$, increases at a
lower rate than $\sigma_0$, hence a decreasing behavior of
$\sigma^*$ is observed. For $\xi=2.04$, a less pronounced, but
also decreasing behavior of $\sigma^*$ is seen. The former
decreasing trend of $\sigma^*$ vs $\gamma$ is reversed for
$\xi=4.08$ and more dramatically for $\xi=16.32$. Curiously, the
four shown curves cross at $\gamma$$\approx$$1.75$, and below this
point,
$\sigma^*(\eta=1.02)<$$\sigma^*(\eta=2.04)<$$\sigma^*(\eta=4.08)<$
$\sigma^*(\eta=16.32)$. This in turn implies that $\sigma^*$ is
dominated by size correlations within this interval. Above this
interval electrostatic correlations dominate. From this
perspective, electrostatic and size correlations do not appear to
be symmetric but, by contrary, are unsymmetric.  Both trends,
increasing and decreasing, of the $\sigma^\ast$ vs $\gamma$ curves
are supported by computer simulations in studies of {CR} in
cylindrical \cite{jimenez_2001} and spherical geometries
\cite{messinaEPL_2002}.

\section{Conclusions}
\label{conclusions}

We studied the electrical double layer (EDL) of a planar wall next
to a restricted primitive model electrolyte through a well
established integral equations theory for inhomogeneous fluids,
i.e., the hypernetted chain/mean spherical approximation
(HNC/MSA). The HNC/MSA equations were derived in terms of the
fundamental parameters of the system: the particles volume
fraction, $\eta$$=$$\frac{\pi}{6}\rho_T a^3$, the ion-ion,
$\xi$$=$$\frac{\beta q^2}{\epsilon a}$, and ion-surface Coulomb
coupling, $\gamma$$=$$2\pi q\beta \sigma_0 a/\epsilon $. To the
best of our knowledge, for the first time a charge reversal
diagram for the planar EDL is formally constructed. In this
diagram two regimes of charge reversal are identified: In regime I
(oscillatory) , charge reversal occurs for any value of $\gamma$
different from zero and independently of the external field
geometry whereas, in regime II (non-oscillatory), a sufficiently
high value of $\gamma$ is required. Using numerical analysis we
found a simple formula which distinguishes these two regimes:
regime I is defined by the condition $\xi \eta \ge 0.0662$,
whereas the regime II is defined by $\xi \eta < 0.0662$.
This simple formula for estimating when charge reversal is
expected is relevant for colloid science, since oscillations of
the charge profiles have been associated with an effective
attractive interaction between like charged macroparticles
\cite{lozada90a,kekicheff95,jimenezJCP06}.
The onset curve of charge reversal for the oscillatory regimes is
in good agreement with the onset points of charge reversal
obtained by molecular simulation for several EDL systems.
Our study pointed out a symmetry in charge reversal between short
range and electrostatic ionic correlations. The results of our
study cover a broad range of practical interest.

\acknowledgments F. J.-A. thanks Y. Duda and G. Odriozola for
their critical reading and comments.

%
%%

%\acknowledgments We thank CONACYT (L007E and C086A) and NEGROMEX.


\begin{thebibliography}{10}
\bibitem{noteCR}
This is not a phase diagram since, from a thermodynamical point of
view, charge
  reversal is not a phase.
\end{thebibliography}
\end{document}